\documentclass[12pt,cite,epsf,epsfig]{article}
\usepackage{epsfig}

\begin{document}
\author{S. Dev\thanks{dev5703@yahoo.com} $^{,1}$,
Sanjeev Kumar\thanks{sanjeev.verma@thapar.edu} $^{,2}$ and Surender
Verma\thanks{ s\_7verma@yahoo.co.in} $^{,1}$}
\title{CP-odd weak basis invariants and texture zeros}
\date{$^1$\textit{Department of Physics, Himachal Pradesh University,\\ Shimla 171005, INDIA.}\\
\smallskip
$^2$\textit{School of Physics and Material Science, Thapar
University,\\ Patiala 147004, INDIA.}}

\maketitle

\begin{abstract}

We construct the CP-odd weak basis invariants from the neutrino mass matrix in a weak basis in
which the charged lepton mass matrix is diagonal and find the
necessary and sufficient conditions for CP conservation. We study the interrelationships between different CP-odd weak basis invariants to examine their implications for the Dirac and Majorana type CP violating phases for the phenomenologically allowed Frampton-Glashow-Marfatia texture zero structures of the neutrino mass matrix.
\end{abstract}

\section{Introduction}

In the Standard Model (SM) of fundamental particles and
interactions, there is no CP violation in the leptonic sector.
However, in most extensions of the SM, there can be several CP
violating phases. In the simplest three generation scenario, there
is a Dirac type CP violating phase in the lepton mixing matrix.
However, for Majorana neutrinos, there could be two additional
phases. It is possible to work in a parametrization in which all
the three CP violating phases are situated in the charged current
lepton mixing matrix. Without any loss of generality, one can work
in flavor basis in which charged lepton mass matrix is diagonal so
that the neutrino mass matrix carries all the information about CP
violation. The search for CP violation in the leptonic sector at
low energies is one of the major challenges for experimental
neutrino physics. Experiments with superbeams and neutrino beams
from neutrino factories have the potential to measure either
directly the Dirac phase $\delta$ through the observation of CP
and T asymmetries or indirectly through neutrino oscillations. An
alternative method is to to measure the area of the unitarity
triangles defined for the leptonic sector. Thus, neutrino physics
provides an invaluable tool for the investigation of leptonic CP
violation at low energies apart from having profound implications
for the physics of the early universe.

In the flavor basis, the mass matrix for Majorana neutrinos
contains nine physical parameters including three mass
eigenvalues, three mixing angles and three CP violating phases.
Two squared mass differences ($\Delta m^2_{12}$ and $\Delta
m^2_{13}$) and two mixing angles ($\theta_{12}$ and $\theta_{23}$)
have been measured in solar, atmospheric and reactor experiments.
The third mixing angle $\theta_{13}$ and the Dirac type CP
violating phase $\delta$ are expected to be measured in the
forthcoming neutrino oscillation experiments. Possible measurement
of effective Majorana mass in neutrinoless double beta decay
searches will provide an additional constraint on the remaining
three neutrino parameters viz. neutrino mass scale and two
Majorana type CP violating phases. While the neutrino mass scale
will be independently determined from the direct beta decay
searches and cosmological observations, the two Majorana phases
will not be uniquely determined from the measurement of effective
Majorana mass even if the overall neutrino mass scale is known.
Thus, it is not possible to fully reconstruct the neutrino mass
matrix from the observations from feasible experiments. Under the
circumstances, it is natural to employ other theoretical inputs
for the reconstruction of the neutrino mass matrix. The possible
forms of these additional theoretical inputs are constrained by
the existing neutrino data. Several proposals have been made in
literature to restrict the possible forms of the neutrino mass
matrix by reducing the number of free parameters which include the
presence of texture zeros \cite{1,2,3}, the requirement of zero
determinant \cite{4}, the zero trace condition \cite{5} amongst
others.

There have been numerous attempts aimed at understanding the
pattern of the fermion masses and mixings by introducing Abelian
and non-Abelian flavor symmetries some of which lead to texture
zeros in the fermion mass matrices. Furthermore, as discussed
earlier, it is not possible to fully reconstruct the neutrino mass
matrix solely from the results of presently feasible experiments
and the introduction of texture zeros is an extra ingredient aimed
at reducing the number of free parameters. However, some sets of
these texture zeros can be obtained by suitable weak basis
transformations and have no physical meaning as such. However, a
large class of sets of leptonic texture zeros considered in the
literature imply the vanishing of certain CP-odd weak basis
invariants and one can, thus, recognize a lepton flavor model in
which the texture zeros are not explicitly present but which
corresponds to a particular texture structure in a certain weak
basis. The presence of texture zeros, in general, leads to a
decrease in the number of independent CP violating phases. A
particular texture zero structure gives rise to definite
relationships between different CP violating phases \cite{2}. Such
exact relations in closed form were obtained in Ref. \cite{6}.
Correlations between Dirac and Majorana CP violating phases for a
particular texture zero scheme were studied in detail in Ref.
\cite{2}. It is, therefore, important to examine the
interrelationships between the CP odd weak basis invariants which
are required to vanish as a necessary and sufficient condition for
CP conservation. It is the purpose of the present work to examine
systematically such interrelationships in terms of the weak basis
invariants constructed from the elements of the neutrino mass
matrix.

\section{Weak basis invariants from the neutrino mass matrix}

The texture zeros are not weak basis (WB) invariants \cite{7}.
This means that a given set of texture zeros which arise in a
certain WB may not be present or may appear in different entries
in another WB. A large class of sets of leptonic texture zeros
considered in the literature imply the vanishing of certain CP-odd
weak-basis invariants \cite{7}. Thus, we can recognize a lepton
mass model in which the texture zeros are not explicitly present
and which corresponds to a particular texture scheme in a certain
WB. The relevance of CP-odd WB invariants in the analysis of the
texture zero ansatze is due to the fact that texture zeros lead to
a decrease in the number of the independent CP violating phases. A
minimum number of CP-odd WB invariants can be found which will all
vanish for the CP invariant mass matrices as a necessary and
sufficient condition \cite{8}.

A necessary and sufficient condition for low energy CP invariance
in the leptonic sector is that the following three WB invariants
are identically zero \cite{9}:
\begin{equation}
I_1=Img~Det[H_{\nu},H_l],
\end{equation}
\begin{equation}
I_2=Img~Tr[H_l M_{\nu}M^*_{\nu}M_{\nu}H^*_l M_{\nu}^*],
\end{equation}
\begin{equation}
I_3=Img~Det[M^*_{\nu}H_l M_{\nu},H^*_l].
\end{equation}
Here, $M_l$ and $M_{\nu}$ are the mass matrices for the charged
leptons and the neutrinos, respectively, and  $H_l=M_l^{\dag}M_l$
and $H_{\nu}=M_{\nu}^{\dag}M_{\nu}$. The invariant $I_1$ was first
proposed by Jarlskog \cite{10} as a rephasing invariant measure of
Dirac type CP violation in the quark sector. It, also, describes
the CP violation in the leptonic sector and is sensitive to the
Dirac type CP violating phase. The invariants $I_2$ and $I_3$ were
proposed by Branco, Lavoura and Rebelo \cite{11} as the WB
invariant measures of Majorana type CP violation. The invariant
$I_3$ was shown \cite{12} to have the special feature of being
sensitive to Majorana type CP violating phase even in the limit of
the exactly degenerate Majorana neutrinos.

The CP violation in the lepton number conserving (LNC) processes
is contained in Jarlskog CP invariant $J$ which can be calculated
from the WB invariant $I_1$ using the relation
\begin{equation}
I_1=-2J(m_e^2-m^2_{\mu})(m^2_{\mu}-m^2_{\tau})(m^2_{\tau}-m^2_e)
(m^2_1-m^2_2)(m^2_2-m^2_3)(m^2_3-m^2_1)
\end{equation}
if the neutrino mass matrix $M_{\nu}$ is a complex symmetric
matrix with eigenvalues $m_1$, $m_2$ and $m_3$ and the charged
lepton mass matrix $M_l$ is diagonal and given by
\begin{equation}
M_l=diag\{m_e,m_{\mu},m_{\tau}\}.
\end{equation}
Then, we have
\begin{equation}
I_1=2(m^2_e-m^2_{\mu})(m^2_{\mu}-m^2_{\tau})(m^2_{\tau}-m^2_e)Img~
(M_{ee}A_{ee}+M_{\mu\mu}A_{\mu\mu}+M_{\tau\tau}A_{\tau\tau})
\end{equation}
where the coefficients $A_{ee}$, $A_{\mu\mu}$ and $A_{\tau\tau}$
are given by
\begin{eqnarray}
A_{ee}=\nonumber M_{\mu\tau}M^*_{e\mu}M^*_{e\tau}
(|M_{\mu\mu}|^2-|M_{\tau\tau}|^2-|M_{e\mu}|^2+|M_{e\tau}|^2)\\
\nonumber +M_{\mu\mu}M^{*2}_{e\mu}(|M_{e\tau}|^2-|M_{\mu\tau}|^2)\\
+M^*_{\mu\mu}M^{*2}_{e\tau}M^2_{\mu\tau},
\end{eqnarray}
\begin{eqnarray}
A_{\mu\mu}=\nonumber M_{e\tau}M^*_{\mu\tau}M^*_{e\mu}
(|M_{\tau\tau}|^2-|M_{ee}|^2-|M_{\mu\tau}|^2+|M_{e\mu}|^2)\\
\nonumber +M_{\tau\tau}M^{*2}_{\mu\tau}(|M_{e\mu}|^2-|M_{e\tau}|^2)\\
+M^*_{\tau\tau}M^{*2}_{e\mu}M^2_{e\tau}
\end{eqnarray}
and
\begin{eqnarray}
A_{\tau\tau}=\nonumber M_{e\mu}M^*_{e\tau}M^*_{\mu\tau}
(|M_{ee}|^2-|M_{\mu\mu}|^2-|M_{e\tau}|^2+|M_{\mu\tau}|^2)\\
\nonumber +M_{ee}M^{*2}_{e\tau}(|M_{\mu\tau}|^2-|M_{e\mu}|^2)\\
+M^*_{ee}M^{*2}_{\mu\tau}M^2_{e\mu}.
\end{eqnarray}
Therefore, the Jarlskog CP invariant measure $J$ is given by
\begin{equation}
J=\frac{Img~(M_{ee}A_{ee}+M_{\mu\mu}A_{\mu\mu}+M_{\tau\tau}A_{\tau\tau})}
{(m^2_1-m^2_2)(m^2_2-m^2_3)(m^2_3-m^2_1)}.
\end{equation}
This relation can be used to calculate $J$ from the mass matrices
directly for any lepton mass model rotated to the WB in which
$M_l$ is diagonal.

The CP violation in lepton number violating (LNV) processes can be
calculated from the WB invariants $I_2$ and $I_3$ which have been
given below:

\begin{eqnarray}
I_2=\nonumber Img~(M_{ee} M_{e \mu}^{*2} M_{\mu \mu}
(m_{e}^2-m_{\mu}^2)^2+  M_{\mu\mu} M_{\mu\tau}^{*2} M_{\tau \tau}
(m_{\mu}^2-m_{\tau}^2)^2+ M_{\tau\tau} M_{e \tau}^{*2} M_{ee} (m_{\tau}^2-m_{e}^2)^2\\
\nonumber +2 M_{ee} M_{e \mu}^* M_{e \tau}^*M_{\mu \tau}
(m_{e}^2-m_{\mu}^2)(m_{e}^2-m_{\tau}^2) \\ \nonumber +2 M_{\mu\mu}
M_{\mu\tau}^* M_{e \mu}^*M_{e \tau}
(m_{\mu}^2-m_{\tau}^2)(m_{\mu}^2-m_{e}^2) \\ +2 M_{\tau\tau} M_{e
\tau}^* M_{\mu \tau}^*M_{e\mu}
(m_{\tau}^2-m_{e}^2)(m_{\tau}^2-m_{\mu}^2))
\end{eqnarray}
and
\begin{equation}
I_3=2(m^2_e-m^2_{\mu})(m^2_{\mu}-m^2_{\tau})(m^2_{\tau}-m^2_e)Img
(m_{e}^2
 M_{ee}B_{ee}+m_{\mu}^2  M_{\mu\mu}B_{\mu\mu}+ m_{\tau}^2  M_{\tau\tau}B_{\tau\tau})
\end{equation}
where the coefficients $B_{ee}$, $B_{\mu\mu}$ and $B_{\tau\tau}$
are given by
\begin{eqnarray}
B_{ee}=\nonumber M_{\mu\tau}M^*_{e\mu}M^*_{e\tau} (m_{\mu}^4
|M_{\mu\mu}|^2-m_{\tau}^4 |M_{\tau\tau}|^2-
m_{e}^2 m_{\mu}^2 |M_{e\mu}|^2+m_{e}^2 m_{\tau}^2 |M_{e\tau}|^2)\\
\nonumber
+m_{\mu}^2 M_{\mu\mu}M^{*2}_{e\mu}(m_{e}^2 |M_{e\tau}|^2-m_{\mu}^2 |M_{\mu\tau}|^2)\\
+m_{\mu}^2 m_{\tau}^2 M^*_{\mu\mu}M^{*2}_{e\tau}M^2_{\mu\tau},
\end{eqnarray}
\begin{eqnarray}
B_{\mu\mu}=\nonumber M_{e\tau}M^*_{\mu\tau}M^*_{e\mu} (m_{\tau}^4
|M_{\tau\tau}|^2-m_{e}^4 |M_{ee}|^2-m_{\mu}^2 m_{\tau}^2
|M_{\mu\tau}|^2+
m_{e}^2 m_{\tau}^2 |M_{e\mu}|^2)\\
\nonumber +m_{\tau}^2 M_{\tau\tau}M^{*2}_{\mu\tau}(m_{\mu}^2 |M_{e\mu}|^2-m_{\tau}^2 |M_{e\tau}|^2)\\
+m_{e}^2 m_{\tau}^2 M^*_{\tau\tau}M^{*2}_{e\mu}M^2_{e\tau}
\end{eqnarray}
and
\begin{eqnarray}
B_{\tau\tau}=\nonumber M_{e\mu}M^*_{e\tau}M^*_{\mu\tau}
(m_{e}^4 |M_{ee}|^2-m_{\mu}^4 |M_{\mu\mu}|^2-m_{e}^2 m_{\tau}^2 |M_{e\tau}|^2+m_{\mu}^2 m_{\tau}^2 |M_{\mu\tau}|^2)\\
\nonumber +m_{e}^2 M_{ee}M^{*2}_{e\tau}(m_{\tau}^2 |M_{\mu\tau}|^2-m_{e}^2 |M_{e\mu}|^2)\\
+m_{e}^2 m_{\mu}^2 M^*_{ee}M^{*2}_{\mu\tau}M^2_{e\mu}.
\end{eqnarray}
From the above expressions for the WB invariants $I_1$, $I_2$ and
$I_3$, one can immediately conclude that

\begin{enumerate}
\item A sufficient condition for CP conservation in LNC and LNV
processes is that all the diagonal entries in the neutrino mass
matrix vanish. Therefore, the symmetric neutrino mass matrices
having Zee-type structure \cite{5} in the diagonal charged lepton
basis are CP conserving.

\item A sufficient condition for CP conservation in LNC processes
is that any three independent entries in the neutrino mass matrix
vanish. Therefore, the symmetric neutrino mass matrices having
three texture zeros in the flavor basis cannot give CP violation
in LNC processes. Such a mass matrix will have three independent
phases and all of them can be rephased away by field
redefinitions. However, some of the symmetric neutrino mass
matrices having three texture zeros in the flavor basis can give
CP violation in LNV processes since $I_2$ can be non-zero. The
neutrino mass matrices having three zeros in a row or a column
will have non-zero $I_2$ and, hence, will exhibit Majorana type CP
violation with one physical phase.

\item The neutrino mass matrix with $\mu-\tau$ symmetry satisfying
the relations
\begin{equation}
M_{e\mu}=\pm ~M_{e\tau} ~and~ M_{\mu\mu}=M_{\tau\tau}
\end{equation}
cannot give CP violation in LNC processes because
$Det~[H_{\nu},H_l]$ vanishes if $\mu-\tau$ symmetry is present.
 However, unlike the LNC
processes, the neutrino mass matrix with $\mu-\tau$ symmetry does
not conserve CP in LNV processes since $I_2$ and $I_3$ do not
vanish, in general, for the neutrino mass matrix with $\mu-\tau$
symmetry.
\end{enumerate}
\section{Implications for texture zeros}

The seven allowed textures of the neutrino mass matrices with
Frampton, Glashow and Marfatia (FGM) texture zero structure have
been summarized in Table 1.

\begin{table}[tb]
\begin{center}
\begin{tabular}{|c|c|}
\hline
 Type  &        Constraining Eqs.         \\
 \hline\hline
 $A_1$ &     $M_{ee}=0$, $M_{e\mu}=0$     \\
 $A_2$ &    $M_{ee}=0$, $M_{e\tau}=0$     \\
 $B_1$ &  $M_{e\tau}=0$, $M_{\mu\mu}=0$   \\
 $B_2$ &  $M_{e\mu}=0$, $M_{\tau\tau}=0$  \\
 $B_3$ &   $M_{e\mu}=0$, $M_{\mu\mu}=0$   \\
 $B_4$ & $M_{e\tau}=0$, $M_{\tau\tau}=0$  \\
 $C$   & $M_{\mu\mu}=0$, $M_{\tau\tau}=0$  \\
  \hline
\end{tabular}
\end{center}
\caption{Allowed two texture zero mass matrices.}
\end{table}

\subsection{Class A}
For neutrino mass matrices of type $A_1$, we have
\begin{eqnarray}
I_1= 2|M_{e\tau}|^2
(m_e^2-m^2_{\mu})(m^2_{\mu}-m^2_{\tau})(m^2_{\tau}-m^2_e)
 Img~(M_{\mu\mu}M_{\tau\tau}M^{*2}_{\mu\tau}),
\end{eqnarray}
\begin{equation}
I_2= (m^2_{\mu}-m^2_{\tau})^2
Img~(M_{\mu\mu}M_{\tau\tau}M^{*2}_{\mu\tau})
\end{equation}
and
\begin{eqnarray}
I_3= 2m_{\mu}^2m_{\tau}^4|M_{e\tau}|^2
(m_e^2-m^2_{\mu})(m^2_{\mu}-m^2_{\tau})(m^2_{\tau}-m^2_e)
Img~(M_{\mu\mu}M_{\tau\tau}M^{*2}_{\mu\tau}).
\end{eqnarray}
The WB invariants for class $A_2$ can be obtained by interchanging
the $\mu$ and $\tau$ indices in the above relations.

An important conclusion follows from the above equations. The CP
violation is brought about by the mismatch in the phases of
$M^2_{\mu\tau}$ and $M_{\mu\mu}M_{\tau\tau}$ whereas the phase of
the element $M_{e\mu}$ or $M_{e\tau}$ plays no role in the CP
violation. In other words, the phase of the element $M_{e\mu}$ or
$M_{e\tau}$ of the neutrino mass matrix has no physical
significance in class A and it can be rephased away. However, a
phase in the $\mu-\tau$ sector of the neutrino mass matrix is
physical and cannot be rephased away. A rephasing transformation
will only rotate the phase in the $\mu-\tau$ sector.

The neutrino mass matrices of class A have four non-zero
independent elements and $J$ is proportional to all of those four
non-zero elements. Any of these four elements cannot vanish since
this will create three texture zeros in the mass matrix making it
incompatible with the existing neutrino data. Hence, a necessary
and sufficient condition for $J=0$ in a complex symmetric neutrino
mass matrix is
\begin{equation}
2arg(M_{\mu\tau})=arg(M_{\mu\mu})+arg(M_{\tau\tau}).
\end{equation}
The condition given in Eq. (20) is equivalent to the condition
that the neutrino mass matrix $M$ can be factorized as $PM^{(r)}P$
where $P$ is a diagonal phase matrix $diag
\{e^{i\phi_1},e^{i\phi_2},e^{i\phi_3}\}$ and $M^{(r)}$ is a real
matrix whose elements are real and equal in magnitude to the
corresponding elements of $M$. Therefore, the necessary and
sufficient condition for CP conservation is that the neutrino mass
matrix $M$ can be written as
\begin{equation}
M=PM^{(r)}P.
\end{equation}
This equation is trivially satisfied for a real mass matrix.
However, if the neutrino mass matrix is a general complex
symmetric matrix, then it is not necessary that it satisfies Eqs.
(20) or (21) and, therefore, it will be CP-violating, in general.

\subsection{Class B}

The neutrino mass matrices of class B can be divided into
sub-classes on the basis of their CP behavior. In the first
sub-class, we have the neutrino mass matrices of type $B_1$ and
$B_2$ for which $I_2$ is zero. In the second sub-class, we have
the neutrino mass matrices of type $B_3$ and $B_4$ for which $I_2$
is non-zero.

For neutrino mass matrices of types $B_1$, we have
\begin{eqnarray}
I_1= 2 (m_e^2-m^2_{\mu})(m^2_{\mu}-m^2_{\tau})(m^2_{\tau}-m^2_e)
Img~(M_{ee}^*M_{\mu\tau}^{*2}M_{e\mu}^2M_{\tau\tau}),
\end{eqnarray}
\begin{equation}
I_2=0
\end{equation}
and
\begin{eqnarray}
I_3=2m_e^2m_{\mu}^2m_{\tau}^2
(m_e^2-m^2_{\mu})(m^2_{\mu}-m^2_{\tau})(m^2_{\tau}-m^2_e)
Img~(M_{ee}^*M_{\mu\tau}^{*2}M_{e\mu}^2M_{\tau\tau}).
\end{eqnarray}
The WB invariants for neutrino mass matrices of type $B_2$ can be
obtained by interchanging the $\mu$ and $\tau$ indices in the
above relations.

For neutrino mass matrices of type $B_3$, we have
\begin{eqnarray}
I_1=2|M_{\mu\tau}|^2
(m_e^2-m^2_{\mu})(m^2_{\mu}-m^2_{\tau})(m^2_{\tau}-m^2_e)
Img~(M_{ee}M_{\tau\tau}M_{e\tau}^{*2}),
\end{eqnarray}
\begin{equation}
I_2=(m^2_{e}-m^2_{\tau}) Img~(M_{ee}M_{\tau\tau}M_{e\tau}^{*2})
\end{equation}
and
\begin{eqnarray}
I_3=2m_{e}^2m_{\tau}^4|M_{\mu\tau}|^2
(m_e^2-m^2_{\mu})(m^2_{\mu}-m^2_{\tau})(m^2_{\tau}-m^2_e)
Img~(M_{ee}M_{\tau\tau}M_{e\tau}^{*2}).
\end{eqnarray}
The WB invariants for class $B_4$ can be obtained from the
interchange of the $\mu$ and $\tau$ indices in the above
relations.

The conditions for the CP invariance of neutrino mass matrices of
class B have been summarized in Table 2. These conditions on the
phases of the neutrino mass matrix can be viewed as the fine
tunings required to have CP invariance.

\subsection{Class C}

For neutrino mass matrices of type $C$, we have
\begin{eqnarray}
\nonumber I_1=2(|M_{e\mu}|^2-|M_{e\tau}|^2)
(m_e^2-m^2_{\mu})(m^2_{\mu}-m^2_{\tau})(m^2_{\tau}-m^2_e)\\
Img~(M_{e\mu}^*M_{e\tau}^*M_{ee}M_{\mu\tau}),
\end{eqnarray}
\begin{eqnarray}
I_2= 2 (m_{e}^2-m_{\mu}^2)(m_{\tau}^2-m_{e}^2)
Img~(M_{e\mu}^*M_{e\tau}^*M_{ee}M_{\mu\tau})
\end{eqnarray}
and
\begin{eqnarray}
\nonumber
I_3=2m_{e}^4(m^2_{\mu}|M_{e\mu}|^2-m^2_{\tau}|M_{e\tau}|^2)
(m_e^2-m^2_{\mu})(m^2_{\mu}-m^2_{\tau})(m^2_{\tau}-m^2_e)\\
Img~(M_{e\mu}^*M_{e\tau}^*M_{ee}M_{\mu\tau})
\end{eqnarray}
The neutrino mass matrix of class C will be CP invariant if the
phases of the mass matrix are fine tuned to satisfy the condition
\begin{equation}
arg(M_{ee})+arg(M_{\mu\tau})=arg(M_{e\mu})+arg(M_{e\tau}).
\end{equation}

\begin{table}[t]
\begin{center}
\begin{tabular}{|c|c|}
\hline
 Type  &       CP invariance condition         \\
 \hline\hline

$A_1$ &  2 $arg(M_{\mu\tau})=arg(M_{\mu\mu})+arg(M_{\tau\tau})$   \\
$B_1$ &  $arg(M_{ee})+2~arg(M_{\mu\tau})=arg(M_{\tau\tau})+2~arg(M_{e\mu})$   \\
$B_3$ &  $2~arg(M_{e\tau})=arg(M_{ee})+arg(M_{\tau\tau})$   \\
$C$ &  $arg(M_{ee})+arg(M_{\mu\tau})=arg(M_{e\mu})+arg(M_{e\tau})$   \\

  \hline
\end{tabular}
\end{center}
\caption{Conditions for CP invariance for the neutrino mass
matrices of class A, B and C.}
\end{table}

The CP violation structure of class C is much different from the
CP violating structure of classes A and B. In class C, the
quantity $I_1$ depends upon the difference between absolute
squares of two matrix elements $M_{e\mu}$ and $M_{e\tau}$ unlike
other classes. So, $I_1$ will vanish if $M_{e\mu}=M_{e\tau}$.
Since, we have $M_{\mu\mu}=M_{\tau\tau}$ in class C, this special
case is $\mu-\tau$ symmetric. Hence, the neutrino mass matrices of
class C having $\mu-\tau$ symmetry can not exhibit CP violation in
LNC processes. However, they can exhibit CP violation in the LNV
processes since $I_2$ and $I_3$ can be non-zero. In this special
case, the neutrino mass matrix contains only a Majorana phase and
there is no Dirac phase. This special case is important, therefore, in the 
sense that we can tell whether the physical phase in the mass matrix 
gives rise to CP violation in LNC or LNV processes.

The three WB invariants for neutrino mass matrices are related
with one another in each class of neutrino mass matrices with FGM
textures. The relations have been summarized in Table 3.
Therefore, $I_1$, $I_2$ and $I_3$ are related with each other and
only one of them is independent. The two Majorana phases
contributing to CP violation in LNV processes will manifest in
$I_2$ and $I_3$ for neutrino mass matrices of types $A_1$, $A_2$,
$B_3$, $B_4$ and $C$. However, they will depend only on the single
invariant $I_3$ for classes $B_1$ and $B_2$. The Dirac phase
contributing to CP violation in LNC processes depends upon $I_1$.
Hence, the three CP violating phases are not independent. There is
only one independent physical phase in the mass matrix and it
contributes to both LNC and LNV processes. However, such a phase
cannot be labelled either Dirac or Majorana unambiguously since it
contributes to the CP violation in both LNC and LNV processes.
Hence, the distinction between Dirac and Majorana phases can not
be maintained in the presence of two texture zeros.  This is contrary to the
conclusions reported in Ref. \cite{13} where it has been claimed that the lone CP violating phase is of Dirac type and there are no Majorana phases.

\begin{table}[tb]
\begin{center}
\begin{tabular}{|c|c|c|}
\hline
 Type  &        $\frac{I_2}{I_1}$   &  $\frac{I_3}{I_1}$     \\
 \hline\hline
 $A_1$ &    $\frac{1}{2 |M_{e\tau}|^2(m_{e}^2-m_{\mu}^2)(m_{\tau}^2-m_{e}^2)}$      &     $m_{\mu}^2m_{\tau}^4$     \\
 $A_2$ &    $\frac{1}{2 |M_{e\mu}|^2(m_{e}^2-m_{\mu}^2)(m_{\tau}^2-m_{e}^2)}$      &     $m_{\mu}^4m_{\tau}^2$     \\
 $B_1$ &  $0$   &   $m_{e}^2m_{\mu}^2m_{\tau}^2$  \\
 $B_2$ &  $0$   &   $m_{e}^2m_{\mu}^2m_{\tau}^2$  \\\
$B_3$ &    $\frac{1}{2 |M_{\mu\tau}|^2(m_{e}^2-m_{\mu}^2)(m_{\mu}^2-m_{\tau}^2)}$      &     $m_{e}^2m_{\tau}^4$     \\
$B_4$ &    $\frac{1}{2 |M_{\mu\tau}|^2(m_{\mu}^2-m_{\tau}^2)(m_{\tau}^2-m_{e}^2)}$      &     $m_{e}^2m_{\mu}^4$     \\
 $C$   & $\frac{1}{(m_{\mu}^2-m_{\tau}^2)(|M_{e\mu}|^2-|M_{e\tau}|^2)}$ & $\frac{m_e^4 (m_{\mu}^2|M_{e\mu}|^2-m_{\tau}^2|M_{e\tau}|^2)}{|M_{e\mu}|^2-|M_{e\tau}|^2}$ \\
 \hline
\end{tabular}
\end{center}
\caption{Ratios $\frac{I_2}{I_1}$ and $\frac{I_3}{I_1}$ for class
A, B and C neutrino mass matrices.}
\end{table}

\section{Conclusions}

We have calculated the CP-odd WB invariants both for LNC and LNV
processes in a basis in which charged lepton mass matrix is
diagonal. These invariants must vanish for CP invariance. We show
that a neutrino mass matrix with three or more zeros gives CP
invariance in LNC processes. However, it can give CP violation
through the WB invariant $I_2$ in LNV processes if the three zeros
are situated along a row of the mass matrix. The neutrino mass
matrices with two texture zeros can, in general, be CP-violating
if they are complex and if their phases are not fine tuned. The
neutrino mass matrices of class C in FGM scheme having $\mu-\tau$
symmetry contain no Dirac phase and show CP violation in LNV
processes only. In general, all the seven classes of the neutrino
mass matrices in FGM texture zero scheme have only one physical
phase which gives rise to CP violation in both LNC and LNV
processes and, hence, cannot be labelled as either Dirac or
Majorana. The physical distinction between a Dirac and a Majorana
phase cannot be maintained in the FGM texture zero scheme.

\end{document}